\documentclass[
reprint,
longbibliography ,
aps,
twoside,
superscriptaddress,
bibnotes
]{revtex4-1}

\usepackage[utf8]{inputenc}
\usepackage{graphicx}
\usepackage{braket}
\usepackage{amsmath, amssymb, mathtools}
\usepackage{bbold}
\usepackage{algorithm, algorithmic}
\usepackage{xcolor}
\usepackage{hyperref}
\usepackage{dsfont}
\usepackage{enumerate}
\usepackage{float}
\usepackage[caption = false]{subfig}

\newlength\figureheight
\newlength\figurewidth
\newcommand{\vac}{\ket{\text{\O}}}
\newcommand{\vacbra}{\bra{\text{\O}}}

\begin{document}

\title{
\textcolor{black}{Protocol for generation of} high-dimensional entanglement from an array of non-interacting photon emitters
}
\author{Thomas J. Bell}
\affiliation{Quantum Engineering Technology Labs, University of Bristol, Bristol BS8 1FD, UK.}
\affiliation{Quantum Engineering Centre for Doctoral Training, University of Bristol, UK.}
\author{Jacob F. F. Bulmer}
\affiliation{Quantum Engineering Technology Labs, University of Bristol, Bristol BS8 1FD, UK.}
\author{Alex E. Jones}
\affiliation{Quantum Engineering Technology Labs, University of Bristol, Bristol BS8 1FD, UK.}
\author{Stefano Paesani}
\affiliation{Center for Hybrid Quantum Networks (Hy-Q), Niels Bohr Institute, University of Copenhagen, Blegdamsvej 17, DK-2100 Copenhagen, Denmark.}
\author{Dara P. S. McCutcheon}
\author{Anthony Laing}
\affiliation{Quantum Engineering Technology Labs, University of Bristol, Bristol BS8 1FD, UK.}
\begin{abstract}
Encoding high-dimensional quantum information into single photons can provide a variety of benefits for quantum technologies, such as improved noise resilience. 
However, the efficient generation of on-demand, high-dimensional entanglement was thought to be out of reach for current and near-future photonic quantum technologies. 
We present a protocol for the near-deterministic generation of $N$-photon, $d$-dimensional photonic \textcolor{black}{Greenberger-Horne-Zeilinger (GHZ)} states using an array of $d$ non-interacting single-photon emitters.
We analyse the impact on performance of common sources of error for quantum emitters, such as photon spectral distinguishability and temporal mismatch, and find they are readily correctable with time-resolved detection to yield high fidelity GHZ states of multiple qudits.
When applied to a quantum key distribution scenario, our protocol exhibits improved loss tolerance and key rates when increasing the dimensionality beyond binary encodings.

\end{abstract}

\maketitle

Encoding quantum information into single photons is a promising route to realising a variety of quantum technologies. 
Most quantum information processing focuses on two-level systems -- \textit{qubit}s -- but for single photons it is possible to encode and process systems of arbitrary dimension -- \textit{qudit}s.
This is exploited for high-dimensional quantum communication~\cite{Islame1701491, nunn2013large, ding2017high, lee2019large}, where moving beyond a binary encoding offers improved bandwidth and noise robustness~\cite{crypt2000Bechmann, Doda2021QuantumEntanglement}. Furthermore, qudits may enable fault-tolerant quantum computation with significantly lower error thresholds~\cite{Enhanced2014Campbell}.

Universal single-qudit gates for photons are possible using efficient interferometer decompositions~\cite{Exp1994Reck, clements2016optimal} and have been demonstrated with high fidelity~\cite{Carolan711}. 
However, generating entanglement between photonic qudits is a challenge for current technologies. Recent work on qudit teleportation introduced probabilistic, linear optical Bell state measurements~\cite{Zhang2019, Luo2019}, but ancillary resources increase and success probabilities decrease as the qudit dimension increases. 
\textcolor{black}{Some solutions to this involve the creation of highly entangled resource states, such as GHZ states, which can be fused to create states which are useful for quantum computation~\cite{bartolucci2021fusion}\cite{Mercedes2015} and communication~\cite{azuma2015all}.
Computation with four-dimensional entangled states has been considered~\cite{Joo2007}, and even qudit Bell pairs (2 photon GHZ states) have been shown to offer advantages in quantum communication over qudit states~\cite{BaccoANetworks}}
All-optical schemes for heralded qudit GHZ state generation were recently shown to be possible~\cite{paesani2021scheme}, but their low success probabilities\textcolor{black}{, which decrease exponentially with qudit dimension and number of photons,} require large overheads to make the entanglement generation near-deterministic.
\textcolor{black}{Experimental progress has been made in creating postselected high-dimensional, multiphoton entanglement~\cite{vigliar2021error}, including 3-photon, 3-dimensional GHZ states~\cite{erhard2018experimental}.
However these postselected schemes are limited in the types of entanglement they can generate~\cite{gu2019quantum}, and because the entanglement is generated upon detection of the photons, they cannot be used for scalable quantum computation.}

In this work, we propose a scheme for the near-deterministic generation of entangled states of photonic qudits.
To produce an $N$-photon, $d$-dimensional GHZ state, we require $d$ quantum emitters, each comprising a spin capable of state-dependent photon emission.
Importantly, our scheme does not require any spin-spin gates.
It takes inspiration from the \textit{photon machine gun} protocol~\cite{Lindner2009} that uses the coherence of a spin-1/2 system with a light-matter interface to deterministically emit entangled photons.
That protocol has been demonstrated in quantum dot systems~\cite{Schwartz434, lee2019quantum} and there is progress towards a realisation using nitrogen vacancy centres in diamond~\cite{vasconcelos2020scalable}. 
There are many error mechanisms in light-matter interfaces that can lead to deterioration of operation or state fidelity~\cite{Kambs_2018}. We investigate how spin dephasing, photon distinguishability, loss, and threshold detection affect the generated photonic qudit entanglement, and show that the requirement of indistinguishable photons can be lifted using time-resolved photon detection, bringing the feasibility of this scheme towards near-term implementation. 

\begin{figure*}
    \centering
    \includegraphics[width=\textwidth, keepaspectratio, clip, trim=0cm 9cm 7.4cm 4cm]{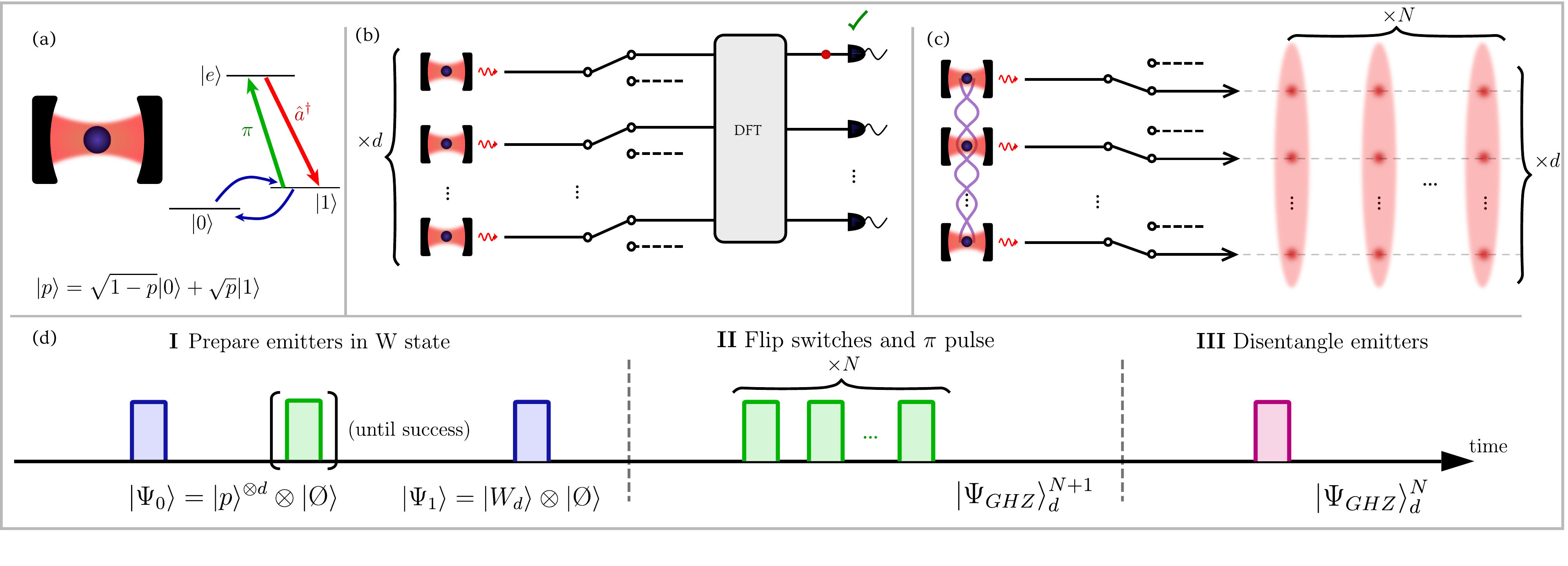}
    \caption{ (a) Level structure for the matter qubit, depicted as an emitter coupled to a cavity. $d$ of these are initialised in state $\ket{p}$. (b) The emitters are $\pi$-pulsed on one transition and emission is routed to a discrete Fourier transform (DFT) interferometer. Detection of one photon (red circle) heralds and projects the emitter array into an entangled W state. This step is repeated until success, at which point (c)~emission is switched away from the DFT interferometer. Sequential $\pi$-pulsing of the emitters then results in emission of a train of $N$ photons, each entangled with one another and the emitters. Each photon \textcolor{black}{(or qudit)} is shown here as an ellipse distributed over $d$ waveguides. Emitters are disentangled via $X$-basis measurements, yielding an $N$ photon qudit GHZ state, \textcolor{black}{of $d$-dimensional qudits}. (d) Steps of the protocol with simplified control pulse sequence \textcolor{black}{on each emitter}, with qubit state control in blue, excitation $\pi$-pulses in green, and qubit measurements in purple. }
    \label{fig::concept}
\end{figure*}

Figure~\ref{fig::concept} shows a platform-agnostic schematic of the idealised setup for generating photonic qudit GHZ states. Figure~\ref{fig::concept}d outlines the \textcolor{black}{required qubit control sequence for each of the} three stages of the state preparation \textcolor{black}{- I. W state preparation in the emitter array, II. Sequential photon emission, III. Disentangling qubit measurements}. We consider the matter system to have three states in a $\Lambda$-configuration with two stable long-lived ground states, only one of which is coupled to a higher level via a radiative transition (see Figure~\ref{fig::concept}a). In the following, the dark and bright ground states correspond to the logical matter qubit states, denoted by $\ket{0}$ and $\ket{1}$ respectively, and the excited state is $\ket{e}$. This level structure enables photon emission conditioned on the state of the spin qubit; application of a $\pi$-pulse and subsequent radiative decay enacts the transformations $\ket{0}\vac \rightarrow \ket{0}\vac$ and $\ket{1}\vac \rightarrow \hat{a}^{\dagger}\ket{1}\vac=\ket{1}\ket{1}$, where $\vac$ denotes vacuum of the electromagnetic field and $\hat{a}^{\dagger}$ is a photon creation operator. This is an effective CNOT gate between the spin qubit and a photonic qubit in the single rail encoding, provided the photonic qubit is initialised in logical 0, the vacuum, as is the case in this protocol.

Each matter system is initialised in a superposition of its ground states, with probability $p$ of being in state $\ket{1}$ (Figure~\ref{fig::concept}a). The total state of the array of $d$ matter systems is
\begin{equation}
\label{eq::initial_state}
    \ket{\Psi_{0}} = \bigotimes_{j=0}^{d-1}\left(\sqrt{1-p}\ket{0}_{j}+ \sqrt{p}\ket{1}_{j}\right)\vac.
\end{equation}
All matter systems are then simultaneously $\pi$-pulsed, and any emitted photons pass to a linear optical network that applies a discrete Fourier transform (DFT) between the modes, erasing the which-path information (Figure~\ref{fig::concept}b). Detection of a single photon in one and only one of the outputs heralds successful generation of a W state over the matter qubit array:
\begin{equation}
    \label{eq::W_state}
    \ket{W_{d}}=\frac{1}{\sqrt{d}}\sum_{j=0}^{d-1}\ket{s_{j}}.
\end{equation}
$\ket{s_{j}}$ is the state in which the $j$th qubit is $\ket{1}$ and all others are $\ket{0}$.
Pulses are repeated until a successful detection event. Note that any detection event will collapse the state of the emitter array, so detection of more than one photon requires re-initialisation of the emitters \textcolor{black}{and is a protocol failure}.



Further $\pi$-pulsing of the emitter array, prepared in $\ket{W_{d}}$, results in emission of a single photon distributed over all $d$ spatial modes (Figure~\ref{fig::concept}c). Labelling the state of a single photon in mode $j$ as $\ket{j}$, this can be interpreted as the superposition over all the basis states of a $d$-dimensional qudit, entangled with the array of matter qubits, $d^{-1/2}\sum _{j=0}^{d-1} \ket{s_{j}}\ket{j}$. 
Repeating this process $N$ times gives a sequence of $N$ time-bin encoded photonic qudits entangled with the matter qubits. After disentangling the emitters by measurement in the $X$-basis, the photons are left in a qudit GHZ state

\begin{equation}
\label{eq::qudit_GHZ}
    \ket{\Psi_{\mathrm{GHZ}}}_{d}^{N} = \frac{1}{\sqrt{d}}\sum_{j=0}^{d-1}(-1)^{m_{j}}\ket{j}^{\otimes N}.
\end{equation}
$m_{j}\in\{0,1\}$ is the measurement outcome for the $j$th emitter.
Any phases incurred due to the particular mode that heralded the W state, and the emitter measurement pattern, can be corrected either by applying phases to individual modes of the output, or by applying rotation gates to the spins. Here and in the following, we have assumed that the detection was at the $0$th detector. 
The time-bin encoding of the protocol may be deterministically converted to any other mode encoding, e.g. to a spatial mode encoding using fast switching and delay lines.

\textbf{Distinguishable photons.} --
A notable difference between this protocol and other machine gun-inspired protocols is the use of multiple emitters, placing additional challenges on the simultaneous control of multiple systems. In practice, non-identical emitters and spectral distinguishability of the emitted photons can degrade the which-path erasure in the DFT, introducing errors that would propagate to the resultant GHZ state. However, interference between different colour photons can be recovered using detectors that resolve inside the photon wavepacket \cite{Legero2003Time-resolvedInterference,Wang2018Time-resolvedColors}. In our scheme, this technique, combined with corrective single-qubit rotations, mitigates spectral distinguishability and enables entanglement generation between quantum emitters. Denoting the spectrum of the photon from the $j$th emitter as $\Phi_{j}(\omega)$, the creation operators for mode $j$ can be re-written $
    \hat{a}_{j}^{\dagger} \rightarrow \int d\omega \Phi_{j}(\omega)\hat{a}_{j}^{\dagger}(\omega)$, 
and the analysis from the ideal case can be repeated. Detection of a single photon at precisely time $t_{0}$ in mode $j$ transforms the state via $\hat{\rho} \rightarrow \hat{E}_{j}^{+}(t_{0}) \hat{\rho} \hat{E}_{j}^{-}(t_{0})/P_{j}(t_{0})$, where $P_{j}$ is a probability density function, and
the electric field operators are Fourier transforms of the spectral operators, $\hat{E}_{j}^{+}(t) = FT\big[\hat{a}_{j}^{\dagger}(\omega)\big]$. Imperfect detector resolution can be treated by convolving this with the detector response function $R(t_{0}, t)$, giving the resultant state of the matter qubits as an incoherent sum over different detection times. After detection of a single photon in the top rail, and no other detection events, the array of qubits is described by the density matrix
\begin{equation}
\label{eq::time_resolved_state}
    \hat{\rho}(t_{0}) = \frac{1}{P(t_{0})}\int dt R(t-t_{0})\sum_{i, j} \zeta_{i}(t)\zeta_{j}^{*}(t)\ket{s_{i}}\bra{s_{j}}.
\end{equation}
The temporal mode functions $\zeta_{j}(t)$ are inverse Fourier transforms of the spectra $\Phi_{j}(\omega)$. The fidelity to the W state $\mathcal{F}_{W} = \bra{W_{d}}\hat{\rho}\ket{W_{d}}$ is then used to evaluate the effect of distinguishable emitters. The consequences for the resultant GHZ state are discussed later.

We consider a Gaussian detector response function and Lorentzian photon lineshapes, given respectively by

\begin{equation}
    \begin{split}
        R(t) &= \frac{1}{\sigma\sqrt{2\pi}} e^{-\frac{t^{2}}{2\sigma^{2}}}, \\
        \Phi_{j}(\omega) &= \sqrt{\frac{\Gamma_{j}}{\pi}} \frac{1}{\Gamma_{j}-i (\omega-\omega_{j})}. \\
    \end{split}
\end{equation}

The parameters $\sigma$, $\omega_{j}$ and $\Gamma_{j}$ have been introduced as the characteristic width of the detector response, characterised by it's jitter, and the central emission frequencies and linewidths of the $j$th emitter.
Substituting these functions into Equation~\ref{eq::time_resolved_state} gives the state of the qubit array. We present some descriptive examples here, and results for the general case in the supplementary information~\cite{SI}.

\begin{figure}[h!]
    \centering
    \subfloat{\includegraphics[scale=1.0, trim = 0 0 -1cm 0, clip]{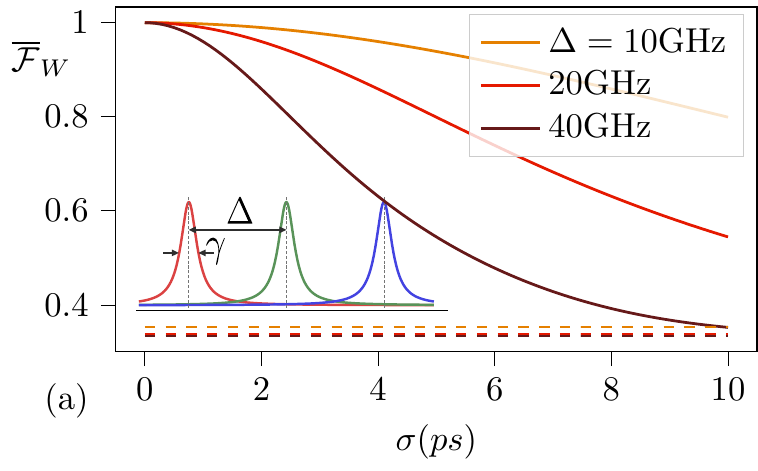}} \\
    \centering
    \subfloat{\includegraphics[scale=1.0]{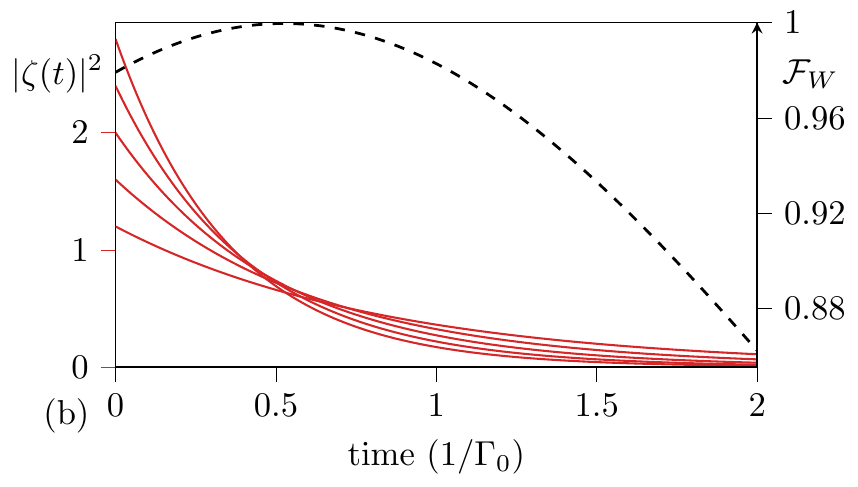}}
    \caption{(a) Heralded W state fidelity as a function of detector resolution for $d=3$ emitters with frequencies $\omega_{0}, \omega_{0} \pm \Delta$, averaged over detection time. Dashed lines show W state fidelity without time resolved detection. Here all emitters have equal linewidths $\Gamma_{j}= 1~\text{GHz}$. (b) Mismatched linewidths: Red curves show probability density for detection of a photon with linewidth [0.6, 0.8, 1, 1.2, 1.4]$\Gamma_{0}$, as a function of time (in units of $1/\Gamma_{0}$), and the corresponding fidelity in black. The maximum fidelity achieved here for detection at $t \sim 0.5$ is 99.9\%, with a time-averaged fidelity of 98\%\textcolor{black}{, obtained using the analytic expression given in the supplementary information~\cite{SI}, and verified with numerical integration.}}
    \label{fig::different_colour}
\end{figure}

We first consider the case of equal linewidths and differening central emission frequencies.
A phase factor oscillating at the frequency difference of each pair of emitters, $\Delta_{j,k}t_{0}$, appears in the relevant term of the density matrix. For a narrow $R(t_{0}, t)$ (small $\sigma$), this phase is well defined, and can be corrected by single qubit rotations on the emitters or phases on the output rails of the eventual GHZ state, so that time-filtering is not required. Phase correction does not fully lift the dependence on detection time, so a useful metric to consider is the time averaged fidelity of the corrected state, which is found to have the simple form $\overline{\mathcal{F}}_{W} = \frac{1}{d^{2}} \sum_{j,k}
e^{-\frac{1}{2}\sigma^{2}\Delta_{j,k}^{2}}$. As seen in Figure~\ref{fig::different_colour}a, high-fidelity path erasure between distinguishable emitters can be achieved when the detector can resolve their frequency difference. A larger $\sigma$ results in a detector that samples a mixture of phases. The phase correction is then imperfect, resulting in a loss of coherence and effective dephasing of the resultant W state, with a dephasing parameter that is a monotonic function of the detector resolution. 3~ps resolution has recently been demonstrated \textcolor{black}{for threshold detectors} \cite{Korzh2020DemonstrationDetector}, corresponding to a correctable frequency splitting of $\sim 50$~GHz\textcolor{black}{, and 16~ps for PNRDs \cite{Zhu2020ResolvingTaper}}. If the detector resolution and emitter beatnotes are long compared to the characteristic photon lifetime, it is not necessary or useful to correct for the additional phase factor. The uncorrected time-averaged fidelity is $\frac{1}{d^{2}} \sum_{j,k} \frac{2\Gamma}{i\Delta_{j,k} + 2\Gamma}$.

\setlength\figureheight{2in}
\setlength\figurewidth{0.45\textwidth}

We also consider the effect of variation in photon linewidths, for example as a result of different cavity leakage rates. Linewidth variation due to differing decay constants leads in general to a `tilted' state with imbalanced amplitudes~\cite{Campbell2007EfficientErasure}. For example, in the $d=2$ case a heralding event at time $t$ yields the state $\lambda_{1}(t)\ket{10} + \lambda_{2}(t)\ket{01}$, with a critical time $t_{c}$ for which $\lambda_{1}(t_{c})=\lambda_{2}(t_{c})$, and we can achieve perfect path erasure~\cite{Campbell2007EfficientErasure}. For higher dimensions the behaviour is analogous, and while generally there will no longer be a critical time that recovers perfect interference, high fidelities are achievable. Figure \ref{fig::different_colour}b shows that up to 99.9\% fidelity can be recovered for a five-dimensional W state with linewidths varying by $\pm40\%$, with a time-average value of 98\%. The highest fidelities are achieved by post-selecting on optimal arrival times ($t_{0}\sim 0.5/\Gamma_{0}$ s in Figure \ref{fig::different_colour}b).

Different photon arrival times may be similarly described in this framework by substituting $\zeta_{j}(t) \rightarrow \zeta_{j}(t-\delta t)$ in Equation \ref{eq::time_resolved_state}. For like emitters, time delays induce both tilting of the resultant W state, as particular emitters are more or less likely to have emitted the detected photon depending on detection time, as well as a relative phase that oscillates at the carrier frequency. It is therefore of critical importance to ensure path length stabilisation in the interferometer as well as highly synchronised pumping of the emitters. 
Integrated photonics, enabling monolithic integration of emitters and circuits~\cite{wan2020}, is an ideal platform to maintain high path length stability when increasing the dimensionality.

\textbf{Loss.} -- Photon loss due to, for example, finite collection efficiency, propagation in waveguides and detector inefficiency is an important source of error in this protocol. We assume equal losses on each mode of the DFT interferometer, with overall photon capture probability per photon denoted $\eta$.  
Loss in the first step of the protocol (Figure~\ref{fig::concept}b) can lead to false heralding events, whereby multiple photons could be emitted and all but one lost prior to detection. This gives the mixed state:

\begin{equation}
\label{eq::noisy_W_state}
    \rho = \alpha_{0}\ket{W_{d}}\bra{W_{d}} + \sum_{\mathcal{Q}} \alpha_{q}\ket{1}\bra{1}_{q}\otimes \ket{W}\bra{W}_{q'}.
\end{equation}
The sum runs over all partitions $\mathcal{Q}$ of the mode indices $[d]$ into two sets $\{q, q'\}$, where $q$ represents the modes where photons have been lost. This corresponds to summing over all possible emission configurations. The partition $q=[\ ]$, $q'=[d]$ has been explicitly separated, as this represents the target W state, where no photons have been lost. The coefficients $\{\alpha_{\mathcal{Q}}\}$ are determined by binomial statistics.

The fidelity is used to determine the quality of W state preparation in the presence of loss. Analytic expressions for $\mathcal{F}_{W}$ and success probability $P_{W}$ are derived in the supplementary information~\cite{SI}, and we find that the fidelity of the heralded state can be made arbitrarily close to unity at the cost of vanishing success probability, by reducing $p$, the weighting of the bright state in the initial superposition. In this manner, losses are converted into time overheads, with a `repeat-until-success' approach. In suitable parameter regimes, the impact of losses can be mitigated using a ``many-successes" scheme. Emitters are $\pi$-pulsed again after a successful generation attempt, without re-initialisation. If there are multiple successive single click events, the erroneous part of the state can be suppressed~\cite{SI}.
Assuming the emitter array has been prepared in an ideal W state, robustness against loss is naturally achieved for protocols which permit postselection. Because the qudit is encoded by a single photon distributed over $d$ waveguides, loss will take the state out of the qudit subspace.
This is a heralded error, corresponding to qudit erasure, and its probability is independent of the qudit dimension.
However, for a noisy W state (Equation~\ref{eq::noisy_W_state}), loss can cause multi-photon emissions to manifest as single photon events, degrading the quality of the resultant GHZ state. The magnitude of this error is determined by $\eta$, and the size of the parameters $\{\alpha_{\mathcal{Q}}\}$. This highlights the importance of preparing a high-fidelity W state at the start of our protocol.
 
The exact parameters of the system under consideration determine the the maximum tolerable number of entanglement attempts and the optimal value of $p$. In practice, this is constrained by the coherence properties of the single photon emitters. Dephasing is a common error for many physical systems, arising from, for example, interaction of an electronic NV spin with nuclear spins of the environment, and its effect is accounted for here by considering a pure-dephasing channel of magnitude $\gamma$ at each time step of the protocol. The relevant Kraus operators commute with protocol operations, as they leave invariant the populations of $\ket{0}$ and $\ket{1}$. Interleaving $k$ of them is thus equivalent to applying a single pure dephasing channel on the resultant state, with $\gamma_{T} = 1 - (1-\gamma)^{k}$. This suppresses off-diagonal terms of the W state by a factor $(1-\gamma_{T})^{2}$, quadratically worse than the dephasing on a single emitter:
 \begin{multline}
    \ket{W_{d}}\bra{W_{d}} \rightarrow \\ (1-\gamma)^{2k}\ket{W_{d}}\bra{W_{d}} + \frac{1-(1-\gamma)^{2k}}{d}\sum_{j=0}^{d-1} \ket{s_{j}}\bra{s_{j}}.
\end{multline}
The entanglement structure of the GHZ state causes this to propagate to a non-local error on the final state, to a degree dictated by the total run-time of the protocol. 
\textcolor{black}{Comparing the magnitude of this error to another possible imperfection - spin relaxation errors, which cause a linear degradation of the state fidelity (as discussed in the supplementary information~\cite{SI}) - we see that dephasing is likely to be the most significant error channel on the qubit array.}

Threshold detectors can operate at much faster rates than photon number-resolving detectors (PNRDs)~\cite{Korzh2020DemonstrationDetector}, but photon bunching can no longer be differentiated from the target single photon outcomes. Similarly to loss, this degrades $\mathcal{F}_{W}$ by introducing additional mixedness to the state in the form of higher photon number terms. Expressions for $\mathcal{F}_{W}$ and $P_{W}$ with threshold detectors can be determined by calculating the size of multi-photon detection events, derived in the supplementary information~\cite{SI}. In order to achieve the same fidelity as PNRDs, $p$ must be reduced and this incurs a significant penalty in the success probability. However, as $d$ is increased, the impact of threshold detectors is lessened, due to the reduced likelihood of photon bunching in the interferometer. For high $d$, success rates approach those achievable using PNRDs. We also consider time-resolved threshold detection with different colour photons in the supplementary information~\cite{SI}, and find that distinguishable emitters reduce photon bunching in the DFT, lessening the detrimental effects of threshold detection -- in fact, the W state fidelity achievable with distinguishable photons and time-resolved threshold detection can exceed that achievable by threshold detectors in the indistinguishable case.

\begin{figure}[h!]
    \centering
    \subfloat{\includegraphics[scale=1.0, trim = 0 0 -1cm 0, clip]{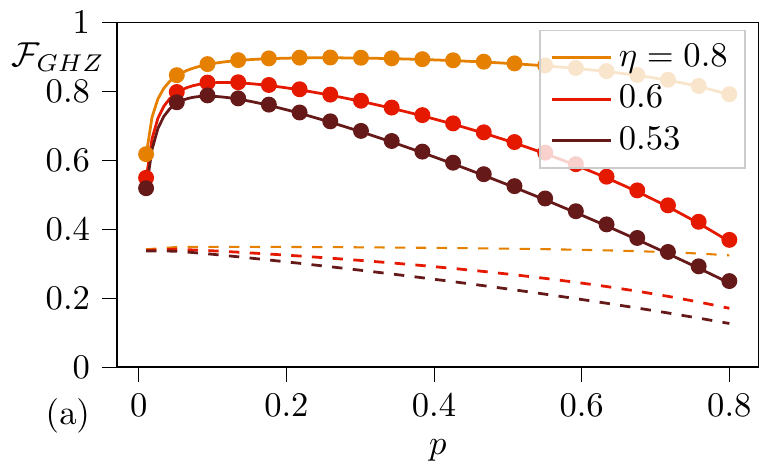}} \\
    \centering
    \subfloat{\includegraphics[scale=1.0, trim = -0.42cm 0 0 0, clip]{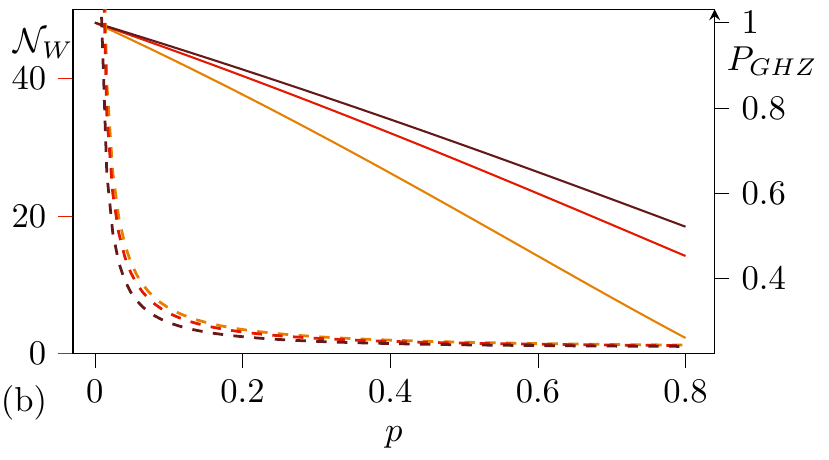}}
    \caption{(a) Fidelity of a 3 qutrit GHZ state generated using emitters with frequencies $[f_{0}$, $f_{0}\pm10$GHz$]$ in this protocol. Solid lines show the fidelity achievable with state of the art 3ps detector resolution, calculated analytically, and dashed lines with no time resolution. Dots are simulated results. Dephasing occurs at a rate 0.0088 per round of emission. The W state is assumed to have been heralded after $1/P_{W}$ entanglement generation attempts. (b) Success probability of the protocol \textcolor{black}{$P_{GHZ}$ (solid lines) and expected number of repetitions until non-vacuum herald $\mathcal{N}_{W}$ (dashed)}. Larger $p$ reduces protocol run-time, but increases the probability of multiple-photon detection events in the repeat-until-success scheme, a failure event which requires re-initialisation.}
    \label{fig::GHZ_fid}
\end{figure}

To calculate the ultimate GHZ state fidelity, we consider each source of error separately, and calculate the resultant fidelity by taking the product $\mathcal{F}_{GHZ} = \mathcal{F}_{dist} \cdot \mathcal{F}_{loss} \cdot \mathcal{F}_{dephase}$. The dephasing-type errors arising from distinguishability and pure-dephasing do not affect the photon emission probability, but reduce its coherence. It can therefore be verified that the degradation of the GHZ-state fidelity is equal to the degradation of W state fidelity prior to the $X$-measurement. The pure-dephasing parameter is determined by considering the mean runtime of the protocol, using analytically calculated success probabilities. Multiplying the dephasing-type errors in this way will lower bound the resultant fidelity, giving good accuracy when one source of dephasing is small. To determine the impact of loss on the final GHZ state we calculate the size of the $k$-excitation terms (when $k$ photons are emitted) in the noisy W state (equation \ref{eq::noisy_W_state}), and find the probability they give rise to single photon events in the second step of the protocol (Figure~\ref{fig::concept}c). The fidelity can be calculated by finding the overlap of each $k$-excitation term with the target GHZ state after $n$ noisy emission events. This approach is verified with numerical simulation. 
\textcolor{black}{The overall protocol success probability $P_{GHZ}$ is determined by the probability of multiple detection events during W-state generation, which constitute an intolerable error requiring re-initialisation of the emitters. $P_{GHZ}$ is the probability that the first non-vacuum detection is a valid heralding event, $P_{GHZ} = P_{W}/(1-P_{vac})$. As shown in Figure \ref{fig::GHZ_fid}(b), this approaches unity for small $p$, but the number of W-state generation attempts required in the repeat-until-success step (and protocol run time) becomes exponentially large as $p\rightarrow0$.}
In Figure~\ref{fig::GHZ_fid}(a), we plot the expected state fidelity when using this protocol to generate a 3-qutrit GHZ state from distinguishable emitters with and without time resolving detection, using realistic emitter parameters -- recent experiments with quantum dot sources~\cite{Tomm2021APhotons} have shown end-to-end efficiencies of 53\% with 76MHz repetition rate and $>1.5\mu s$ dephasing time, a corresponding dephasing rate of $\gamma<0.0088$ per photon. We observe that 80\% fidelity is achievable in this regime, with success probability of $>95\%$, which can be boosted to $\mathcal{F}_{GHZ} > 90\%$ with high probability for modest improvements to capture efficiency and dephasing time. \textcolor{black}{Recent developments in detector timing resolution~\cite{Korzh2020DemonstrationDetector, Zhu2020ResolvingTaper} and efficiency~\cite{Akhlaghi2015, Reddy20} raise the prospect of simultaneously fast and efficient detectors in the near future.}

In the supplementary information~\cite{SI}, we calculate secure key rates that could be achieved using these figures of merit for qudit Bell state generation. We follow the methods in \cite{BaccoANetworks,Doda2021QuantumEntanglement}, and find that five such emitters could be used to distribute $d=5$ qudit Bell pairs with secure bit rates of 1.3~Mbps [using the optimal value of $p=0.056$]. Higher $d$ results in recoverable key rates where losses are prohibitive in the lower-dimensional case.

\textbf{Discussion.} We have shown in this work that highly entangled photonic qudit states can be generated with rate overheads that scale promisingly with $d$, even between non-identical sources. Practically, significant work must be done to simultaneously control and synchronise multiple emitters, but through the use of time resolving detection we have shown that the effects of distinguishability from non-identical emitters can be mitigated without time-filtering, significantly relaxing this challenging experimental requirement. Looking ahead, one could consider entanglement distillation protocols with a broker-client scheme, as in~\cite{Campbell2008Measurement-basedLoss} to boost success probabilities from other herald patterns.

\begin{acknowledgments}
\textit{Acknowledgements} - TJB acknowledges support from UK EPSRC (EP/SO23607/1); Fellowship  support  from  EPSRC  is acknowledged by AL (EP/N003470/1). 
We acknowledge support from the EPSRC Hub in Quantum Computing and Simulation (EP/T001062/1) and Networked Quantum Information Technologies (EP/N509711/1).
\end{acknowledgments}

\bibliographystyle{vancouver}

\bibliography{refs.bib}

\begin{thebibliography}{10}

\bibitem{Islame1701491}
Islam NT, Lim CCW, Cahall C, Kim J, Gauthier DJ.
\newblock Provably secure and high-rate quantum key distribution with time-bin
  qudits.
\newblock Science Advances. 2017;3(11).
\newblock Available from:
  \url{https://advances.sciencemag.org/content/3/11/e1701491}.

\bibitem{nunn2013large}
Nunn J, Wright L, S{\"o}ller C, Zhang L, Walmsley I, Smith B.
\newblock Large-alphabet time-frequency entangled quantum key distribution by
  means of time-to-frequency conversion.
\newblock Optics express. 2013;21(13):15959-73.

\bibitem{ding2017high}
Ding Y, Bacco D, Dalgaard K, Cai X, Zhou X, Rottwitt K, et~al.
\newblock High-dimensional quantum key distribution based on multicore fiber
  using silicon photonic integrated circuits.
\newblock npj Quantum Information. 2017;3(1):1-7.

\bibitem{lee2019large}
Lee C, Bunandar D, Zhang Z, Steinbrecher GR, Dixon PB, Wong FN, et~al.
\newblock Large-alphabet encoding for higher-rate quantum key distribution.
\newblock Optics express. 2019;27(13):17539-49.

\bibitem{crypt2000Bechmann}
Bechmann-Pasquinucci H, Tittel W.
\newblock Quantum cryptography using larger alphabets.
\newblock Phys Rev A. 2000 May;61:062308.
\newblock Available from:
  \url{https://link.aps.org/doi/10.1103/PhysRevA.61.062308}.

\bibitem{Doda2021QuantumEntanglement}
Doda M, Huber M, Murta G, Pivoluska M, Plesch M, Vlachou C.
\newblock {Quantum Key Distribution Overcoming Extreme Noise: Simultaneous
  Subspace Coding Using High-Dimensional Entanglement}.
\newblock Physical Review Applied. 2021 3;15(3):034003.
\newblock Available from:
  \url{https://link.aps.org/doi/10.1103/PhysRevApplied.15.034003}.

\bibitem{Enhanced2014Campbell}
Campbell ET.
\newblock Enhanced Fault-Tolerant Quantum Computing in $d$-Level Systems.
\newblock Phys Rev Lett. 2014 Dec;113:230501.
\newblock Available from:
  \url{https://link.aps.org/doi/10.1103/PhysRevLett.113.230501}.

\bibitem{Exp1994Reck}
Reck M, Zeilinger A, Bernstein HJ, Bertani P.
\newblock Experimental realization of any discrete unitary operator.
\newblock Phys Rev Lett. 1994 Jul;73:58-61.
\newblock Available from:
  \url{https://link.aps.org/doi/10.1103/PhysRevLett.73.58}.

\bibitem{clements2016optimal}
Clements WR, Humphreys PC, Metcalf BJ, Kolthammer WS, Walmsley IA.
\newblock Optimal design for universal multiport interferometers.
\newblock Optica. 2016;3(12):1460-5.

\bibitem{Carolan711}
Carolan J, Harrold C, Sparrow C, Mart{\'\i}n-L{\'o}pez E, Russell NJ,
  Silverstone JW, et~al.
\newblock Universal linear optics.
\newblock Science. 2015;349(6249):711-6.
\newblock Available from:
  \url{https://science.sciencemag.org/content/349/6249/711}.

\bibitem{Zhang2019}
Zhang C, Chen JF, Cui C, Dowling JP, Ou ZY, Byrnes T.
\newblock Quantum teleportation of photonic qudits using linear optics.
\newblock Phys Rev A. 2019 Sep;100:032330.
\newblock Available from:
  \url{https://link.aps.org/doi/10.1103/PhysRevA.100.032330}.

\bibitem{Luo2019}
Luo YH, Zhong HS, Erhard M, Wang XL, Peng LC, Krenn M, et~al.
\newblock Quantum Teleportation in High Dimensions.
\newblock Phys Rev Lett. 2019 Aug;123:070505.
\newblock Available from:
  \url{https://link.aps.org/doi/10.1103/PhysRevLett.123.070505}.

\bibitem{bartolucci2021fusion}
Bartolucci S, Birchall P, Bombin H, Cable H, Dawson C, Gimeno-Segovia M, et~al.
\newblock Fusion-based quantum computation.
\newblock arXiv preprint arXiv:210109310. 2021.

\bibitem{Mercedes2015}
Gimeno-Segovia M, Shadbolt P, Browne DE, Rudolph T.
\newblock From Three-Photon Greenberger-Horne-Zeilinger States to Ballistic
  Universal Quantum Computation.
\newblock Phys Rev Lett. 2015 Jul;115:020502.
\newblock Available from:
  \url{https://link.aps.org/doi/10.1103/PhysRevLett.115.020502}.

\bibitem{azuma2015all}
Azuma K, Tamaki K, Lo HK.
\newblock All-photonic quantum repeaters.
\newblock Nature communications. 2015;6(1):1-7.

\bibitem{Joo2007}
Joo J, Knight PL, O'Brien JL, Rudolph T.
\newblock One-way quantum computation with four-dimensional photonic qudits.
\newblock Phys Rev A. 2007 Nov;76:052326.
\newblock Available from:
  \url{https://link.aps.org/doi/10.1103/PhysRevA.76.052326}.

\bibitem{BaccoANetworks}
Bacco D, Bulmer JFF, Erhard M, Huber M, Paesani S.
\newblock Proposal for practical multidimensional quantum networks.
\newblock Phys Rev A. 2021 Nov;104:052618.
\newblock Available from:
  \url{https://link.aps.org/doi/10.1103/PhysRevA.104.052618}.

\bibitem{paesani2021scheme}
Paesani S, Bulmer JFF, Jones AE, Santagati R, Laing A.
\newblock Scheme for Universal High-Dimensional Quantum Computation with Linear
  Optics.
\newblock Phys Rev Lett. 2021 Jun;126:230504.
\newblock Available from:
  \url{https://link.aps.org/doi/10.1103/PhysRevLett.126.230504}.

\bibitem{vigliar2021error}
Vigliar C, Paesani S, Ding Y, Adcock JC, Wang J, Morley-Short S, et~al.
\newblock Error-protected qubits in a silicon photonic chip.
\newblock Nature Physics. 2021;17(10):1137-43.

\bibitem{erhard2018experimental}
Erhard M, Malik M, Krenn M, Zeilinger A.
\newblock Experimental greenberger--horne--zeilinger entanglement beyond
  qubits.
\newblock Nature Photonics. 2018;12(12):759-64.

\bibitem{gu2019quantum}
Gu X, Erhard M, Zeilinger A, Krenn M.
\newblock Quantum experiments and graphs II: Quantum interference, computation,
  and state generation.
\newblock Proceedings of the National Academy of Sciences.
  2019;116(10):4147-55.

\bibitem{Lindner2009}
Lindner NH, Rudolph T.
\newblock Proposal for Pulsed On-Demand Sources of Photonic Cluster State
  Strings.
\newblock Phys Rev Lett. 2009 Sep;103:113602.
\newblock Available from:
  \url{https://link.aps.org/doi/10.1103/PhysRevLett.103.113602}.

\bibitem{Schwartz434}
Schwartz I, Cogan D, Schmidgall ER, Don Y, Gantz L, Kenneth O, et~al.
\newblock Deterministic generation of a cluster state of entangled photons.
\newblock Science. 2016;354(6311):434-7.
\newblock Available from:
  \url{https://science.sciencemag.org/content/354/6311/434}.

\bibitem{lee2019quantum}
Lee J, Villa B, Bennett A, Stevenson R, Ellis D, Farrer I, et~al.
\newblock A quantum dot as a source of time-bin entangled multi-photon states.
\newblock Quantum Science and Technology. 2019;4(2):025011.

\bibitem{vasconcelos2020scalable}
Vasconcelos R, Reisenbauer S, Salter C, Wachter G, Wirtitsch D, Schmiedmayer J,
  et~al.
\newblock Scalable spin--photon entanglement by time-to-polarization
  conversion.
\newblock npj Quantum Information. 2020;6(1):1-5.

\bibitem{Kambs_2018}
Kambs B, Becher C.
\newblock Limitations on the indistinguishability of photons from remote solid
  state sources.
\newblock New Journal of Physics. 2018 nov;20(11):115003.
\newblock Available from: \url{https://doi.org/10.1088/1367-2630/aaea99}.

\bibitem{Legero2003Time-resolvedInterference}
Legero T, Wilk T, Kuhn A, Rempe G.
\newblock {Time-resolved two-photon quantum interference}.
\newblock In: Applied Physics B: Lasers and Optics. vol.~77. Springer; 2003. p.
  797-802.
\newblock Available from:
  \url{https://link.springer.com/article/10.1007/s00340-003-1337-x}.

\bibitem{Wang2018Time-resolvedColors}
Wang XJ, Jing B, Sun PF, Yang CW, Yu Y, Tamma V, et~al.. {Time-resolved boson
  sampling with photons of different colors}. arXiv; 2018.
\newblock Available from:
  \url{https://journals.aps.org/prl/abstract/10.1103/PhysRevLett.121.080501}.

\bibitem{SI}
see Supplementary Information;.

\bibitem{Korzh2020DemonstrationDetector}
Korzh B, Zhao QY, Allmaras JP, Frasca S, Autry TM, Bersin EA, et~al.
\newblock {Demonstration of sub-3 ps temporal resolution with a superconducting
  nanowire single-photon detector}.
\newblock Nature Photonics. 2020 4;14(4):250-5.
\newblock Available from: \url{https://doi.org/10.1038/s41566-020-0589-x}.

\bibitem{Zhu2020ResolvingTaper}
Zhu D, Colangelo M, Chen C, Korzh BA, Wong FNC, Shaw MD, et~al.
\newblock Resolving Photon Numbers Using a Superconducting Nanowire with
  Impedance-Matching Taper.
\newblock Nano Letters. 2020;20(5):3858-63.
\newblock PMID: 32271591.
\newblock Available from: \url{https://doi.org/10.1021/acs.nanolett.0c00985}.

\bibitem{Campbell2007EfficientErasure}
Campbell ET, Fitzsimons J, Benjamin SC, Kok P.
\newblock {Efficient growth of complex graph states via imperfect path
  erasure}.
\newblock New Journal of Physics. 2007 6;9(6):196.
\newblock Available from: \url{http://www.njp.org/}.

\bibitem{wan2020}
Wan NH, Lu TJ, Chen KC, Walsh MP, Trusheim ME, De~Santis L, et~al.
\newblock Large-scale integration of artificial atoms in hybrid photonic
  circuits.
\newblock Nature. 2020;583(7815):226-31.

\bibitem{Tomm2021APhotons}
Tomm N, Javadi A, Antoniadis NO, Najer D, L{\"{o}}bl MC, Korsch AR, et~al.
\newblock {A bright and fast source of coherent single photons}.
\newblock Nature Nanotechnology. 2021 1:1-5.
\newblock Available from: \url{https://doi.org/10.1038/s41565-020-00831-x}.

\bibitem{Akhlaghi2015}
Akhlaghi M, Schelew E, Young J.
\newblock Waveguide integrated superconducting single-photon detectors
  implemented as near-perfect absorbers of coherent radiation.
\newblock Nature communications. 2015;6(1).

\bibitem{Reddy20}
Reddy DV, Nerem RR, Nam SW, Mirin RP, Verma VB.
\newblock Superconducting nanowire single-photon detectors with 98\% system
  detection efficiency at 1550nm.
\newblock Optica. 2020 Dec;7(12):1649-53.
\newblock Available from:
  \url{http://www.osapublishing.org/optica/abstract.cfm?URI=optica-7-12-1649}.

\bibitem{Campbell2008Measurement-basedLoss}
Campbell ET, Benjamin SC.
\newblock {Measurement-based entanglement under conditions of extreme photon
  loss}.
\newblock Physical Review Letters. 2008 9;101(13):130502.
\newblock Available from:
  \url{https://journals.aps.org/prl/abstract/10.1103/PhysRevLett.101.130502}.

\end{thebibliography}

\newpage

\onecolumngrid
\section{Supplementary Information}
\subsection{General detection patterns}
Here we show that detection events in general modes result in local correctable phases in W state preparation
Starting from Equation \ref{eq::initial_state} of the main text, after preparing each emitter in the superposition state $\ket{p}$, $\pi$-pulsing results in the state
\begin{equation}
    \bigotimes^{d-1}_{j=0}\left[\sqrt{1-p}\ket{0}_{j}+ \sqrt{p}\hat{a}_{j}^{\dagger}\ket{1}_{j}\right]\vac.
\end{equation}
After interference on a $d$-mode DFT, the state of the system is
\begin{equation}
\label{eq::DFT}
    \ket{\Psi} = \bigotimes^{d-1}_{j=0}\left[\sqrt{1-p}\ket{0}_{j}+ \sqrt{\frac{p}{d}} \sum^{d-1}_{k=0}\exp{\left(i \frac{2\pi jk}{d}\right)}\hat{a}_{k}^{\dagger} \ket{1}_{j}\right]\vac.
\end{equation}
Detection of a single photon in mode $l$ then projects the emitters into the state
\begin{equation}
\begin{split}
     \vacbra\hat{a}_{l}\ket{\Psi} =& \vacbra\hat{a}_{l} \bigotimes^{d-1}_{j=0}\left[\sqrt{1-p}\ket{0}_{j}+ \sqrt{\frac{p}{d}} \sum^{d-1}_{k=0}\exp{\left(i \frac{2\pi jk}{d}\right)}\hat{a}_{k}^{\dagger}\ket{1}_{j}\right]\vac \\
     =& \vacbra\hat{a}_{l} \bigotimes^{d-1}_{j=0}\left[\sqrt{1-p}\ket{0}_{j}+ \sqrt{\frac{p}{d}} \exp{\left( i \frac{2\pi jl}{d} \right)}\hat{a}_{l}^{\dagger}\ket{1}_{j}\right]\vac \\
     =& \sum_{j=0}^{d-1} (1-p)^{\frac{d-1}{2}}\left(\frac{p}{d}\right)^{\frac{1}{2}} \exp{\left( i \frac{2\pi jl}{d} \right)} \ket{s_{j}}
\end{split}
\end{equation}
where the state $\ket{s_{j}}$ represents the $j$th emitter in $\ket{1}$ and all others in $\ket{0}$. This state then is the W state, with phases $\frac{2\pi jl}{d}$ on the $j$th term.

\subsection{W state fidelity and generation probability in the presence of loss}

 The effect of loss on a single mode is treated by inserting variable reflectivity beamsplitters between target modes and ancillas, and tracing out of the ancilla modes, equivalent to an amplitude damping channel. For the initial state described in the main text, the effect of loss leads to a density matrix for the matter-photon system:

\begin{equation}
\begin{split}
\label{eq::loss_deriv1}
    \rho = \bigotimes_{j}  \left[\ket{\psi_{0}}_{j}\bra{\psi_{0}}_{j} + \ket{\psi_{1}}_{j}\bra{\psi_{1}}_{j}\right]&, \\
    \ket{\psi_{0}}_{j} = \left(\sqrt{1-p}\ket{0}_{j} + \sqrt{\eta}\sqrt{p}\hat{a}^{\dagger}_{j}\ket{1}_{j}\right)\vac,  \qquad
    \ket{\psi_{1}}_{j} &= \sqrt{1-\eta}\sqrt{p}\ket{1}_{j}\vac.
\end{split}
\end{equation}
The states $\ket{\psi_{0/1}}_{j}$ correspond to the resultant states if a photon has or has not been lost from mode $j$. The DFT transforms surviving creation operators as before, causing $\ket{\psi_{0}}_{j}$ to assume the form of Equation \ref{eq::DFT}.
Defining operators $\hat{A}_{j}$, $\hat{B}_{j}$ such that $\ket{\psi_{0}}_{j} = \hat{A}_{j}\vac, \ket{\psi_{1}}_{j} = \hat{B}_{j}\vac$, the resultant light-matter density matrix can be written succinctly as 
\begin{equation}
\label{eq::state}
    \rho = \bigotimes_{j=0}^{d-1}\left[\hat{A}_{j}\vac\vacbra\hat{A}_{j}^{\dagger} + \hat{B}_{j}\vac\vacbra\hat{B}_{j}^{\dagger}\right] 
    = \sum_{s\in \{0,1\}^{d}} \mathbf{\hat{A}}^{(\mathbf{1}\oplus s)}\hat{\mathbf{B}}^{s} \vac\vacbra \mathbf{\hat{A}^{\dagger^{(\mathbf{1}\oplus s)}}}\mathbf{\hat{B}^{\dagger^{s}}}
\end{equation}
where $\mathbf{A}^{s} = \hat{A}_{0}^{s_{0}}\hat{A}_{1}^{s_{1}}...\hat{A}_{d-1}^{s_{d-1}}$, $\mathbf{1}$ represents the all 1s string, and $\oplus$ denotes binary addition. The sum runs over all binary strings of length $d$, and string $s$ here represents the set of photons that have been lost.
We can now consider post-selection on a single click in the top rail. The resultant reduced density operator for the matter system is $\mathcal{N}\text{Tr}\left[\hat{a}_{0}^{\dagger}\vac\vacbra\hat{a}_{0}\rho\right]$
where the trace is performed over photonic degrees of freedom, and $\mathcal{N}$ is a normalisation factor. The $s=\mathbf{0}$ string gives the target W-state, multiplied by $\eta$. The $s=\mathbf{1}$ string does not contribute as it is a vacuum state of the field. All the other strings will contribute, and act to reduce the fidelity of the final state with the target W-state. The magnitude of this state is determined by the number of 1s in the string, referred to as the distance $\delta$. For a particular string with distance $\delta_{s}$, the inner product gives:
\begin{equation}
    \vacbra\hat{a}_{0}\mathbf{\hat{A}}^{(\mathbf{1}\oplus s)}\hat{\mathbf{B}}^{s} \vac = (1-\eta)^{\frac{\delta_{s}}{2}}p^{\frac{\delta_{s}}{2}}(1-p)^{\frac{d-\delta_{s}-1}{2}}\sqrt{\frac{p\eta(d-\delta_{s})}{d}}\ket{\chi(s)} 
\end{equation}
Where the states $\ket{\chi(s)}$ are the superposition of all states where $\delta_{s}$ 1s are fixed by the string s, and there is a single additional 1. The magnitude of this term is determined by binomial statistics, calculated according to the number of configurations for this event to happen. For example if $s$ is the distance 1 string (100...0), then the state is
\begin{equation}
        \ket{\chi(s)} = \frac{1}{\sqrt{d-1}}\left(\ket{110..0} + \ket{101...0} + ... + \ket{100...1}\right) = \ket{1}_{0}\otimes\ket{W}_{d-1}
\end{equation}

There will be $ {d \choose \delta_{s}}$ unique strings of distance $\delta_{s}$. The final reduced density matrix for the matter states takes the form
\begin{equation}
    \rho_{m} = \frac{\eta p(1-p)^{d-1}\ket{W_{d}}\bra{W_{d}} + \sum_{s\notin\{\mathbf{0},\mathbf{1}\}}\left|\alpha(\delta_{s})\right|^{2}\ket{\chi(s)}\bra{\chi(s)})}{\sum_{s\neq \mathbf{1}}|\alpha(\delta_{s})|^{2}},
\end{equation}
where the term corresponding to the W state has been explicitly separated. The probability of detecting this state, and the corresponding fidelity are then simply calculated as $P = \sum_{s\neq \mathbf{1}}|\alpha(\delta_{s})|^{2}$ and $\mathcal{F}_{W} = |\alpha(0)|^{2}/P$. As discussed in the previous section, there are $d$ possible `good' detection events - one for each output mode, so the total success probability $P_{W} = Pd$ and so
\begin{equation}
\label{eq::w_fid_loss}
    \mathcal{F}_{W} = \frac{\eta p(1-p)^{d-1}}{ \sum_{k=0}^{d-1}\frac{d!}{\delta !(d-k)!}\left|\alpha(\delta)\right|^{2}} = \left(\sum_{k=0}^{d-1} \frac{(d-1)!}{\delta !(d-\delta-1)!} p^{\delta}(1-p)^{-\delta}(1-\eta)^{\delta}\right)^{-1}, 
\end{equation}

\begin{equation}
    P_{W} = \sum_{\delta=0}^{d-1} \frac{d!}{\delta !(d-\delta-1)!} p^{\delta+1}(1-p)^{d-\delta-1}\eta(1-\eta)^{\delta}.
\end{equation}
These expressions are verified by simulation.

\subsubsection{\textcolor{black}{Amplitude damping}}
\textcolor{black}{
In some platforms, particularly those with non-degenerate computational basis states of the emitter, relaxation of the qubit can be a relevant error mechanism. This is treated straightforwardly by inserting an amplitude damping channel after each round of pumping, with Kraus operators $\{E_{k}\} = \{\ket{0}\bra{0} + \sqrt{1-\lambda}\ket{1}\bra{1}$, $ \sqrt{\lambda}\ket{0}\bra{1}\}$ acting on each qubit, where $\lambda$ gives the magnitude of the error. The effect of this is similar to an effective loss that increases with the protocol runtime. Before heralding of the W-state, the state of the system after amplitude damping is 
\begin{equation}
    \begin{split}
    \rho = \bigotimes_{j}  \left[\ket{\phi_{0}}_{j}\bra{\phi_{0}}_{j} + \ket{\phi_{1}}_{j}\bra{\phi_{1}}_{j}\right]&, \\
    \ket{\phi_{0}}_{j} = \left(\sqrt{1-p}\ket{0}_{j} + \sqrt{1-\lambda}\sqrt{p}\hat{a}^{\dagger}_{j}\ket{1}_{j}\right)\vac,  \qquad
    \ket{\phi_{1}}_{j} &= \sqrt{\lambda}\sqrt{p}\ket{0}_{j}\vac.
\end{split}
\end{equation}
This equation takes a similar form to Equation \ref{eq::loss_deriv1}, and we see that the heralded state on a single click is now
\begin{equation}
\label{eq::noisy_W_state_ad}
    \rho = \beta_{0}\ket{W_{d}}\bra{W_{d}} + \sum_{\mathcal{Q}} \beta_{q}\ket{0}\bra{0}_{q}\otimes \ket{W}\bra{W}_{q'},
\end{equation} where the sum runs over all partitions of the $d$ emitters into two sets, and the coefficients $\beta_{q}$ can be calculated in the same way as for loss above, making the substitution $\eta\rightarrow1-\lambda$. The subsequent degradation of W state fidelity due to this error mode is similarly given by making this substitution in equation \ref{eq::w_fid_loss}. After heralding the W-state, the effect of this channel can be investigated by applying it to the perfect W state. We find 
\begin{equation}
    \ket{W_{d}}\bra{W_{d}} \rightarrow (1-\lambda)\ket{W_{d}}\bra{W_{d}} + \lambda\ket{\mathbf{0}}\bra{\mathbf{0}},
\end{equation}
with $\ket{\mathbf{0}}$ representing the all-zero state. The fidelity scales linearly with the the degree of damping. This linear scaling is less detrimental that the quadratic error associated with dephasing errors discussed in the main text. Furthermore, in post-selectable protocols it can be detected and removed, as it is outside the qudit computational space. Combining this with the performance metrics of currently viable single photon emitters, we deem pure-dephasing to be the dominant error mechanism of concern acting on the emitters.
}
\subsubsection{Thresholding}
To include threshold detection, the detection projector must be modified, $\hat{a}_{j}^{\dagger}\vac\vacbra\hat{a}_{j} \rightarrow (\hat{\mathds{1}}_{j} - \vac_{j}\vacbra_{j}) \bigotimes_{k\neq j}\vac_{k}\vacbra_{k}$, which represents any non-zero number of excitations in mode $j$ and vacuum in all other modes. We again consider a click in the 0 mode, for simplicity.

\begin{equation}
    \rho = \mathcal{N} \sum_{m=1}^{d}\frac{1}{m!}\sum_{s\in \{0,1\}^{d}}
    \text{Tr}\left[(\hat{a}_{0}^{\dagger})^{m}\vac\vacbra\hat{a}^{m}_{0}
    \mathbf{\hat{A}}^{(\mathbf{1}\oplus s)}\hat{\mathbf{B}}^{s}
    \vac\vacbra \mathbf{\hat{A}^{\dagger^{(\mathbf{1}\oplus s)}}}\mathbf{\hat{B}^{\dagger^{s}}}\right]
\end{equation}

For a given string $s$, we can explicitly partition the state into $U(s)$ and $U'(s)$, where $u \in U(s)$ if $s_{u} = 0$. The number of indices in $U(s)$ is then $|U(s)| = d-\delta_{s}$. After projecting onto the vacuum for all other modes, the projection onto the $m$-photon state is
\begin{equation}
\begin{split}
\frac{1}{\sqrt{m!}}\bra{\text{\O}}_{0}\hat{a}_{0}^{m} \bigotimes_{u, u'}\left[\left(\sqrt{1-p}\ket{0}_{u} + \sqrt{\frac{p\eta}{d}}\hat{a}^{\dagger}_{0}\ket{1}_{u}\right)\sqrt{p(1-\eta)}\ket{1}_{u'}\right] \vac_{0} \\
= \frac{1}{\sqrt{m!}}\bra{\text{\O}}_{0}\hat{a}_{0}^{m} \sum_{\xi \in\{0,1\}^{d-\delta_{s}}} \sqrt{\frac{p\eta}{d}}^{\delta_{\xi}} \sqrt{1-p}^{d-\delta_{s}-\delta_{\xi}} \ket{\xi} (\hat{a}_{0}^{\dagger})^{\delta_{\xi}}\otimes(\sqrt{p(1-\eta)})^{\delta_{s}}\ket{1}_{U'} \vac_{0} \\
= \sum_{\xi_{m}}\sqrt{m!(1-p)^{d-\delta_{s}-m}\left(\frac{p\eta}{d}\right)^{m}(p(1-\eta))^{\delta_{s}}}\ket{\xi_{m}}\otimes\ket{1}_{U'}
\end{split}
\end{equation}
where $\xi_{m}$ are the strings of distance $m$, corresponding to $m$ photon incidences. The heralded state is then found by summing over all strings $s$ and $\xi$,
\begin{equation}
    \rho = \mathcal{N} \sum_{{\substack{s \in \{0,1\}^{d} \\ s\neq \mathbf{1}}}} \sum_{m=1}^{d} m!(1-p)^{d-\delta_{s}-m}\left(\frac{p\eta}{d}\right)^{m}(p(1-\eta))^{\delta_{s}} \sum_{{\substack{\xi\in\{0,1\}^{d-\delta_{s}} \\ \delta_{\xi}=m}}} \ket{\xi}_{U}\bra{\xi}_{U} \otimes \ket{1}_{U'}\bra{1}_{U'}
\end{equation}

from which we can calculate the total detection probability
\begin{equation}
    P_{W} = d\sum_{\delta_{s}=0}^{d-1} {d\choose \delta_{s}}  \sum_{\delta_{\xi}=1}^{d-\delta_{s}} {d-ks \choose \delta_{\xi}} \delta_{\xi}!(1-p)^{d-\delta_{s}-\delta_{\xi}}\left(\frac{p\eta}{d}\right)^{\delta_{\xi}}(p(1-\eta))^{\delta_{s}}
\end{equation}
The index $\delta_{\xi}$ represents the number of photons that caused a detection event, so the $\delta_{s}=0$, $\delta_{\xi}$ term corresponds to the W state, and all others are orthogonal. Thus the fidelity can be calculated as $P_{W}(\delta_{s}=0, \delta_{\xi}=1)/P_{W}$
\begin{equation}
    \mathcal{F}_{W} = \left(\sum_{\delta_{s}=0}^{d-1} \sum_{\delta_{\xi} = 1}^{d-\delta_{s}} \frac{d!}{\delta_{s}!(d-\delta_{s}-\delta_{\xi})!} p^{\delta_{s}+\delta_{\xi}-1}\eta^{\delta_{\xi}-1}d^{-\delta_{\xi}}(1-p)^{1-\delta_{s}-\delta_{\xi}} (1-\eta)^{\delta_{s}}\right)^{-1}
\end{equation}
The case of PNRDs can be recovered by restricting to $\delta_{\xi}=1$.

\subsubsection{Multiple successes}

\setlength\figureheight{2.5in}
\setlength\figurewidth{0.45\textwidth}
\begin{figure}[h!]
\centering
\subfloat{\includegraphics[]{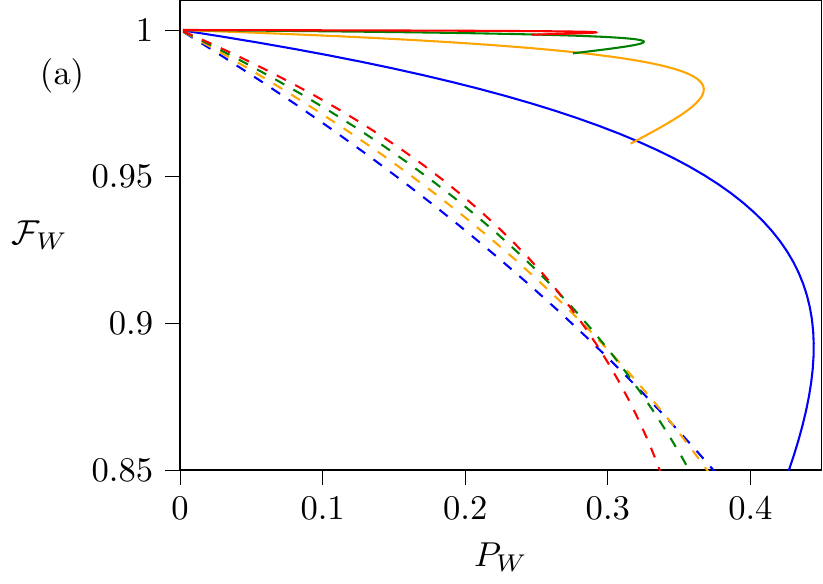}}
\subfloat{\includegraphics[]{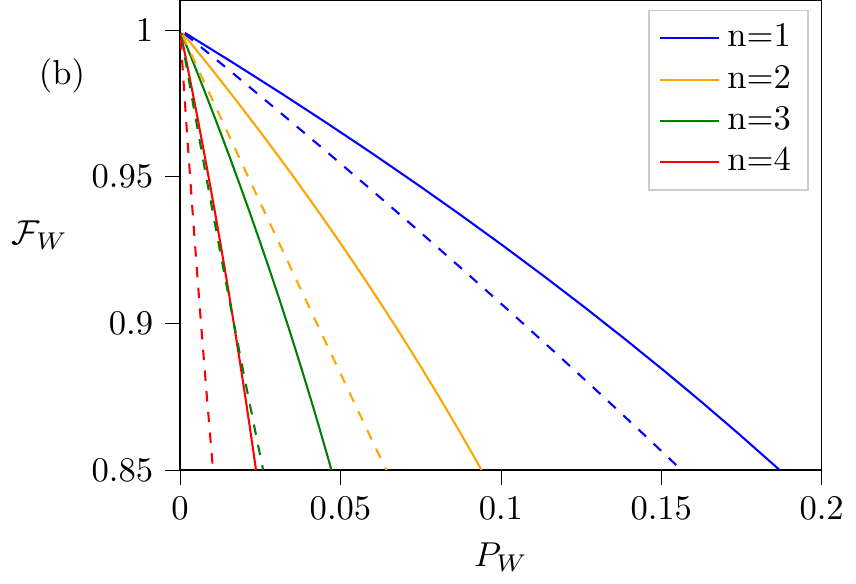}} \\
\subfloat{\includegraphics[]{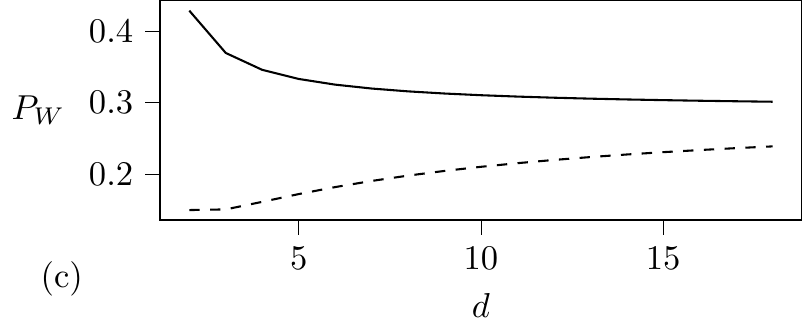}}
\subfloat{\includegraphics[]{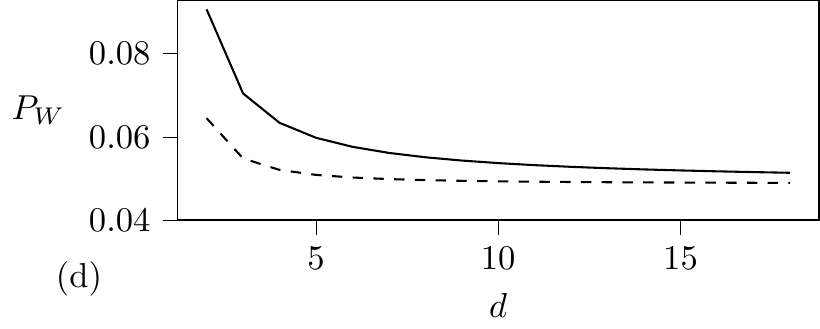}}
\caption{Analytically calculated fidelities $\mathcal{F}_{W}$ and success probabilities $R$ are shown for $n$ successive single click events, with (a) $\eta_{1}=0.9$ and (b) $\eta_{1}=0.5$, for $d=3$ W state preparation. The initial superposition state of the emitters is varied along the curves, with $p\rightarrow0$ in the top left resulting in high fidelity but low success probability. Solid lines show the performance of PNRDs, and dashed lines indicate threshold detectors. In (a), many-successes can give higher quality W states for the same success rates, whereas in (b) greater losses cause this approach to fail. Threshold detectors impose fixed errors and reduce the efficacy of this scheme, but as the qudit dimension is increased, the impact of threshold detectors is lessened. (c) and (d) present the probability of a single click event with $\eta_{1}=0.9$ and $\eta_{1}=0.5$, where $p$ is adjusted to herald a 95\% fidelity W state. As $d$ is increased, the performance penalty for threshold detection diminishes.}
\label{fig::losses}
\end{figure}
\setlength\figureheight{2.5in}

As discussed in the main text, by repeating the excitation procedure after a successful outcome, detection of a second successful outcome can increase the fidelity of the heralded state. To quantify the fidelity after a second iteration, we consider the probabilities that different terms in the heralded state give rise to a single click. This probability is dependent only on the total number of emitters in the bright state $\ket{1}$, which is equal to $\nu = \delta_{s}+\delta_{\xi}$. In order for a $\nu$-photon emission to result in a single click, up to $\nu-1$ photons can be lost, and all remaining photons must end up in the same output port (with PNRDs the remaining number of photons must be 1). This event has probability 
\begin{equation}
    P_{next} = d\sum_{y=1}^{\nu} {\nu \choose y} \frac{y!}{d^{y}}\eta^{y}(1-\eta)^{\nu-y}
\end{equation}

so the general expression for the probability of $n$ rounds of successive single click detection events, using threshold detectors, is:
\begin{equation}
P_{W}(n) = d^{n}\sum_{\delta_{s}=0}^{d-1} \sum_{\delta_{\xi} = 1}^{d-\delta_{s}} \left( \frac{d!}{\delta_{s}!(d-\delta_{s}-\delta_{\xi})!} \left(\frac{p\eta}{d}\right)^{\delta_{\xi}}(1-p)^{d-\delta_{s}-\delta_{\xi}} \left(p(1-\eta)\right)^{\delta_{s}}\Bigg[\sum_{y=1}^{\delta_{s}+\delta_{\xi}}  \frac{(\delta_{s}+\delta_{\xi})!}{(\delta_{s}+\delta_{\xi}-y)!}\frac{\eta^{y}}{d^{y}}(1-\eta)^{\delta_{s}+\delta_{\xi}-y}\Bigg]^{n-1}\right)
\end{equation}
and the resultant fidelity $\mathcal{F}_{W}(n) = dp(1-p)^{d-1}\eta^{n}/P_{W}(n)$. These functions are verified with numerical simulation, and the resulting performance is shown in Figure \ref{fig::losses}, where it is seen that for suitably low losses, adjusting the initial superposition state $\ket{p}$ of the emitters and heralding on multiple successes can greatly improve $\mathcal{F}_{W}$ for a fixed success probability. Threshold detection imposes higher necessary capture probabilities for this approach to be useful, due to the fixed overhead they impose. This expression can also be used to determine the qudit generation probability in the first stage of the protocol. If $n$ successive clicks are used to generate the W state, the probability of heralding a single photon in a qudit emission attempt is $P_{q} = P_{W}(n+1)/P_{W}(n)$. In general, the capture probability during photonic GHZ emission may differ from W state generation. This is easily accounted for by substituting $\eta\rightarrow\eta_{2}$ in $P_{next}$.


\subsection{Time resolved detection}
Here we derive the results for the heralded state obtained via interferometery of distinguishable emitters with a DFT followed by time resolved detection of the state. After interferometery, the matter-photon state is, as before (Equation~\ref{eq::DFT}):
\begin{equation}
    \ket{\Psi} = \bigotimes^{d-1}_{j=0}\left[\sqrt{1-p}\ket{0}_{j}+ \sqrt{\frac{p}{d}} \sum^{d-1}_{k=0}\exp{\left(i \frac{2\pi jk}{d}\right)}\hat{a}_{k}^{\dagger} \ket{1}_{j}\right]\vac.
\end{equation}

We now make the replacements as outlined in the main text
\begin{equation}
    \hat{a}_{j}^{\dagger} \rightarrow \int d\omega \Phi_{j}(\omega)\hat{a}_{j}^{\dagger}(\omega), 
\end{equation}

\begin{equation}
    \ket{\Psi} = \prod^{d-1}_{j=0}\left[\sqrt{1-p}\ket{0}_{j}+ \sqrt{\frac{p}{d}} \sum^{d-1}_{k=0}\exp{\left( i \frac{2\pi jk}{d}\right)}\int d\omega \Phi_{j}(\omega)\hat{a}_{k}^{\dagger}(\omega)\ket{1}_{j}\right]\vac.
\end{equation}
 We introduce the measurement operator $\hat{M}_{i}(t) = \vac\vacbra\hat{E}_{i}^{(+)}(t)$, representing detection of a single photon in mode $i$ and no other detection events. The post-measurement state is found by applying the measurement operator and performing a biased trace over time, where the bias function is the detector response function, representing our knowledge or ignorance about when the photon was detected
 \begin{equation}
     \rho \longrightarrow \frac{1}{P(t)}\int_{-\infty}^{\infty} dt' R(t-t') \hat{M}(t') \ket{\Psi}\bra{\Psi} \hat{M}^{\dagger}(t').
 \end{equation}

 \begin{equation}
 \begin{split}
    \hat{M}(t') \ket{\Psi}
     &= \vac\vacbra\frac{1}{\sqrt{2\pi}} \int d\omega 'e^{-i\omega 't}\hat{a}_{0}(\omega ')  \prod^{d-1}_{j=0}\left[\sqrt{1-p}\ket{0}_{j}+ \sqrt{\frac{p}{d}} \sum^{d-1}_{k=0}\exp{\left( i \frac{2\pi jk}{d} \right)}\int d\omega \Phi_{j}(\omega)\hat{a}_{k}^{\dagger}(\omega)\ket{1}_{j}\right]\vac \\
     &= \sqrt{\frac{(1-p)^{d-1}p}{d}} \sum_{j=0}^{d-1} \zeta_{j}(t') \ket{s_{j}}
     \end{split}
 \end{equation}
where we have used the usual conventions of a click in the 0th mode, and $\zeta_{j}(t)$ is introduced as the temporal mode function of the photon from the $j$th emitter. Without losses, we can restrict to the $d\times d$ single excitation subspace. In this subspace, the state can be described by the $d\times d$ matrix, 
\begin{equation}
    \hat{\rho} = \frac{B}{\text{Tr}\left[B\right]}, \qquad B = \frac{(1-p)^{d-1}p}{d} \sum_{j,k} \int_{-\infty}^{\infty} dt' R(t-t') \zeta_{j}(t') \zeta_{k}^{*}(t') \ket{s_{j}} \bra{s_{k}}
\end{equation}

It remains to evaluate these integrals for some simple cases. This can be done numerically, or analytically. Using the Lorentzian lineshape and Gaussian response function described in the main text, we seek to compute the integral:
\begin{equation}
\begin{split}
    & \int_{-\infty}^{\infty} dt' \frac{1}{\sigma\sqrt{2\pi}} e^{-\frac{(t-t')^{2}}{2\sigma^{2}}} 2\sqrt{\gamma_{j}\gamma_{k}}e^{-i(\omega_{j}-\omega_{k})t'}e^{-(\gamma_{j}+\gamma_{k})t'} \Theta(t').
    \end{split}
\end{equation}
$\Delta_{j,k}$ has been introduced as the frequency difference between a pair of emitters $j, k$.

The final expression for the $j, k$ matrix element is then
\begin{equation}
    B_{j,k}=\frac{(1-p)^{d-1}p}{d} \sqrt{\gamma_{j}\gamma_{k}}   \Gamma_{c}\left( \frac{1}{2\sqrt{2}\sigma}(2(\gamma_{j}+\gamma_{k})\sigma^{2} - 2t + 2i\sigma^{2}\Delta_{j,k})\right)
    e^{\frac{1}{2}(\gamma_{j}+\gamma_{k})^{2}\sigma^{2}} e^{i(\gamma_{j}+\gamma_{k})\sigma^{2}\Delta_{j,k}} e^{-\sigma^{2}\Delta_{j,k}^{2}/2} e^{-i\Delta_{j,k}t} e^{-(\gamma_{j}+\gamma_{k})t}
\end{equation}
This is normalised by its trace. With matching linewidths:
\begin{equation}
    \rho_{j,k} = \frac{1}{d}\frac{\Gamma_{c}\left(\frac{2\gamma \sigma^{2} - t_{0} + i\sigma^{2}\Delta_{j,k}}{\sqrt{2} \sigma}\right)}{\Gamma_{c}\left(\frac{2\gamma \sigma^{2} - t_{0}}{\sqrt{2} \sigma}\right)} e^{2i\gamma\sigma^{2}\Delta_{j,k}}e^{-\frac{1}{2}\sigma^{2}\Delta_{j,k}^2}e^{-i\Delta_{j,k}t} \ket{s_{j}}\bra{s_{k}}
\end{equation}

The fidelity to the W state in this basis is simply calculated by $\mathcal{F}_{W} = \frac{1}{d}\sum_{j,k}\rho_{j,k}$

\subsection{Time averaged fidelity}
The main advantage of time-resolved detection in this scheme is leveraging the detection-time information to apply phase corrections, so that time filtering is not required. Despite this, fidelities do vary with detection time, so a useful operational metric to consider is the average fidelity after heralding on a single click. To calculate this, compute:
\begin{equation}
    \overline{\mathcal{F}}_{W} = \frac{\int_{-\infty}^{\infty} dt \mathcal{F}_{W}(t) P_{W}(t)}{\int_{-\infty}^{\infty} dt P_{W}(t)}.
\end{equation}

We expect the normalisation $\int_{-\infty}^{\infty} dt P_{W}(t) = p(1-p)^{d-1}$, the total probability of a click occurring at any time. For general photon wavepacket and detector response:
\begin{equation}
    \rho_{j,k}(t)P_{W}(t) = \frac{(1-p)^{d-1}p}{d} \int_{-\infty}^{\infty} dt' R(t-t') \zeta_{j}(t') \zeta_{k}^{*}(t') 
\end{equation}
then
\begin{equation}
    \overline{\mathcal{F}}_{W} = \frac{1}{d^{2}} \sum_{j,k} \int_{-\infty}^{\infty}dt \int_{-\infty}^{\infty} dt' R(t-t') \zeta_{j}(t') \zeta_{k}^{*}(t').
\end{equation}
Note that this corresponds to the situation where we have not applied a phase correction based on the click time. In this instance, taking this time average is analogous to tracing out time completely - thus the weighted tracing we have done before becomes irrelevant. Integrating over a normalized response function ($R(t-t')$ is the only term with $t$-dependence), the $t$ integral evaluates to 1 and we are left with
\begin{equation}
    \overline{\mathcal{F}}_{W} = \frac{1}{d^{2}} \sum_{j,k} \int_{-\infty}^{\infty} dt' \zeta_{j}(t') \zeta_{k}^{*}(t').
\end{equation}
This tells us that if you have time resolved detection but don't do anything conditioned on the click time (or forget what it was), on average you may as well have not used time-resolving detectors.
For the characteristic functions described before, this quantity will evaluate to
\begin{equation}
\begin{split}
        \overline{\mathcal{F}}_{W} &= \frac{1}{d^{2}} \sum_{j,k} \int_{0}^{\infty} dt' 2\sqrt{\gamma_{j}\gamma_{k}}e^{-(\gamma_{j}+\gamma_{k})t'} e^{-i\Delta_{j,k}t'} \\
        &= \frac{1}{d^{2}} \sum_{j,k} \frac{2\sqrt{\gamma_{j}\gamma_{k}}}{i\Delta_{j,k} + (\gamma_{j} + \gamma_{k})}.
\end{split}
\end{equation}
This is notably independent of the detector width (as expected). This tells us the fidelity we get if we use a non time-resolving detector.
When performing time corrections, we simply introduce a $e^{i\Delta_{j,k}t}$ term to the heralded density matrix before integration. The time-averaged fidelity will then be:
\begin{equation}
\begin{split}
    \overline{\mathcal{F}}_{W} &= \frac{1}{d^{2}} \sum_{j,k} \int_{-\infty}^{\infty}dt \int_{-\infty}^{\infty} dt' R(t-t') \zeta_{j}(t') \zeta_{k}^{*}(t') e^{i\Delta_{j,k}t} \\
    &= \frac{\sqrt{2\gamma_{j}\gamma_{k}}}{\sigma d^{2}\sqrt{\pi}} \sum_{j,k}\int_{-\infty}^{\infty}dt \int_{0}^{\infty} dt' \exp{\left( -\frac{(t-t')^{2}}{2\sigma^{2}} - i\Delta_{j,k}t'  -(\gamma_{j}+\gamma_{k})t' + i\Delta_{j,k}t \right)} \\ 
         &= \frac{1}{d^{2}} \sum_{j,k}
    \frac{2\sqrt{\gamma_{j}\gamma_{k}}}{\gamma_{j}+\gamma_{k}}
    e^{-\frac{1}{2}\sigma^{2}\Delta_{j,k}^{2}}.
\end{split}
\end{equation}

\subsection{Time resolved threshold detection}
Here we will calculate the fidelity $\mathcal{F}_{W}$ of the state heralded by threshold detectors that can time-resolve within the photon wavepacket.
When combining threshold detection with time resolution, a detection event at time $t$ in mode 0 corresponds to the operation

\begin{equation}
\label{eq::thresholding_time_resolved}
\begin{split}
    \rho \longrightarrow & \vacbra\hat{E}_{0}^{+}(t) \rho \hat{E}_{0}^{-}(t) + \int_{0}^{T} dt_{1}\hat{E}_{0}^{+}(t+t_{1}) \hat{E}_{0}^{+}(t) \rho \hat{E}_{0}^{-}(t) \hat{E}_{0}^{-}(t+t_{1}) \\ &+
    \int_{0}^{T}\int_{0}^{T} dt_{1}dt_{2} \hat{E}_{0}^{+}(t+t_{2})\hat{E}_{0}^{+}(t+t_{1})\hat{E}_{0}^{+}(t)\rho \hat{E}_{0}^{-}(t+t_{2})\hat{E}_{0}^{-}(t+t_{1})\hat{E}_{0}^{-}(t) + ...\vac
    \end{split}
\end{equation}
where the integral is done over the dead time of the detector, which is taken to be much longer that the photon coherence time and can be safely extended to infinity. Field operators are normal ordered, such that the later time electric field operator is to the left(right) for the positive(negative) frequency components. The sum is truncated at the expected maximum possible number of photons in the system. This state is projected on the vacuum by waiting for anything else to happen and postselecting on nothing happening.

After preparing the emitters in the superposition state $\ket{p}$, excitation and interference enacts the transformation
\begin{equation}
\begin{split}
         \ket{\Psi}_{i} &= \prod^{d-1}_{j=0}\left(\sqrt{1-p}\ket{0}_{j}+ \sqrt{p}\int d\omega \Phi_{j}(\omega)\hat{a}_{j}^{\dagger}(\omega) \ket{1}_{j}\right)\ket{\text{\O}} \\ &
        \longrightarrow \prod^{d-1}_{j=0}\left[\sqrt{1-p}\ket{0}_{j}+ \sqrt{\frac{p}{d}} \sum^{d-1}_{k=0}\exp{\left( i \frac{2\pi jk}{d} \right)}\int d\omega \Phi_{j}(\omega)\hat{a}_{k}^{\dagger}(\omega)\ket{1}_{j}\right]\ket{\text{\O}}
\end{split}
\end{equation}
where we have now considered the spectral lineshape $\Phi_{j}(\omega)$ of the emitters. Removing a single quantum from the field in mode 0, and projecting onto vacuum in all other modes results in the conditional (un-normalised) state
\begin{equation}
\begin{split}
    \hat{E}_{0}^{+}(t) \ket{\Psi}_{i} &= \frac{1}{\sqrt{2\pi}} \int d\omega 'e^{-i\omega 't}\hat{a}_{0}(\omega ')  \prod^{d-1}_{j=0}\left[\sqrt{1-p}\ket{0}_{j}+ \sqrt{\frac{p}{d}} \int d\omega \Phi_{j}(\omega)\hat{a}_{0}^{\dagger}(\omega)\ket{1}_{j}\right]\ket{\text{\O}}  \\
    &= \frac{1}{\sqrt{2\pi}} \int d\omega 'e^{-i\omega 't}\hat{a}_{0}(\omega ') \sum_{s \in \{0,1\}^{d}} \sqrt{\frac{p^{\delta_{s}}(1-p)^{d-\delta_{s}}}{d^{\delta_{s}}}} \prod_{j, s_{j}=1} \int d\omega_{j} \Phi_{j}(\omega_{j}) \hat{a}_{j}^{\dagger}(\omega_{j})    \vac \ket{s}\\
    &= \frac{1}{\sqrt{2\pi}}  \sum_{s \in \{0,1\}^{d}} \sqrt{\frac{p^{\delta_{s}}(1-p)^{d-\delta_{s}}}{d^{\delta_{s}}}}  \sum_{i, s_{i}=1} \int d\omega'  e^{-i\omega 't} (\omega_{i}) \Phi_{i}(\omega_{i}) \delta(\omega'-\omega_{i}) \prod_{j \neq i, s_{j}=1} \int d\omega_{j} \Phi_{j}(\omega_{j}) \hat{a}_{j}^{\dagger}(\omega_{j})    \vac \ket{s}\\
    &=  \sum_{s \in \{0,1\}^{d}} \sqrt{\frac{p^{\delta_{s}}(1-p)^{d-\delta_{s}}}{d^{\delta_{s}}}} \sum_{i, s_{i} = 1} \zeta_{i}(t) \prod_{j \neq i, s_{j}=1} \int d\omega_{j} \Phi_{j}(\omega_{j}) \hat{a}_{j}^{\dagger}(\omega_{j})    \vac \ket{s}.
\end{split}
\end{equation}
In the second line the product runs over indices $j$ such that $s_{j}=1$, and in the third line we sum over all states where one of these indices has been picked out by the electric field operator.
The 1-photon term is found by directly projecting this onto the vacuum, which restricts to the single excitation term (those strings with distance 1).
\begin{equation}
    \vacbra\hat{E}_{0}^{+}(t) \ket{\Psi}_{i} = \sqrt{\frac{p(1-p)^{d-1}}{d}} \sum_{i=0}^{d-1} \zeta_{i}(t) \ket{s_{i}}
\end{equation}
and has an overlap with the W-state
\begin{equation}
   \bra{W_{d}}\rho_{1}\ket{W_{d}} = \frac{p(1-p)^{d-1}}{d} \sum_{i, j} \zeta_{i}(t)\zeta_{j}^{*}(t).  
\end{equation}

Higher photon number terms require repeated applications of the electric field operator. We see that the above result can be extended to

\begin{equation}
    \vacbra\hat{E}_{0}^{+}(t+\tau_{n-1})...\hat{E}_{0}^{+}(t)\ket{\Psi}_{i}= \sqrt{\frac{p^{n}(1-p)^{d-n}}{d^{n}}} \sum_{i_{0}...i_{n-1} = 0}^{d-1} \zeta_{i_{0}}(t)...\zeta_{i_{n-1}}(t)  \ket{i_{0}...i_{n-1}}
\end{equation}
where here the state $\ket{i_{0}...i_{n-1}}$ corresponds to $1$s at indices $i$ and $0$ elsewhere. Calculating the fidelity with the W state then requires calculating the size of each terms contribution to the final density matrix - only the one photon term will overlap with the target W state:
\begin{equation}
    \mathcal{F}_{W} = \frac{\bra{W_{d}}\rho_{1}\ket{W_{d}}}{\sum_{n=1}^{d} Tr[\rho_{n}]}.
\end{equation}

\begin{figure}[h!]
    \centering
    \includegraphics[]{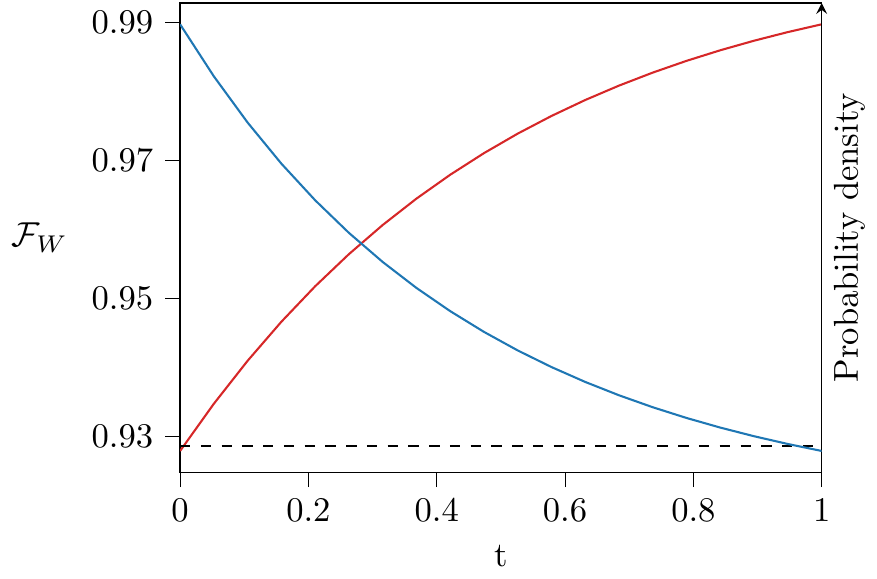}
    \caption{The click probability $P_{W}$ (blue) and fidelity $\mathcal{F}_{W}$ (red) are plotted here for Lorentzian photons detected with a precisely time-resolved threshold detector at time $t$ (in units of $1/\Gamma$), for 3 emitters with frequencies $\omega_{0}$, $\omega_{0} \pm 10\Gamma$. The dashed black line shows the fidelity achievable with the same initial superposition state $\ket{p}^{\otimes 3}$ using threshold detectors with indistinguishable emitters.}
    \label{fig::t_res_threshold}
\end{figure}

The trace of the $n$-photon term is
\begin{equation}
    Tr[\hat{\rho}_{n}] = \frac{p^{n}(1-p)^{d-n}}{d^{n}}\idotsint dt_{1}...dt_{n-1}\left( \sum_{A \in [d]^{n}} \Big| \sum_{\mathcal{P}(A)} \prod_{i=1}^{n} \zeta_{p_{i}}(t_{i}) \Big| ^{2}\right)
\end{equation}
Where we sum over all configurations $A$ of $n$-photon emissions, and for each emission configuration we must consider all permutations of arrival times $\mathcal{P}$. $p_{i}$ is the $i$th element of the permutation $\mathcal{P}$. These integrals are performed numerically, and it is observed that distinguishable emitters and time-resolved detection when thresholding can result in higher fidelity heralded W states than for indistinguishable emitters, as shown in Figure \ref{fig::t_res_threshold}. Intuitively, this is because distinguishability leads to reduced bunching in the outputs of the interferometer, diminishing the detrimental effects of threshold detectors. The other effect observed is the increase in $\mathcal{F}_{W}$ with detection time, due to the $e^{-2n\Gamma t}$ term present in the probability of an $n$-photon detection event occurring at time $t$. 

\subsection{Bit rates}
\begin{figure*}[h!]
\centering
\subfloat{\includegraphics[]{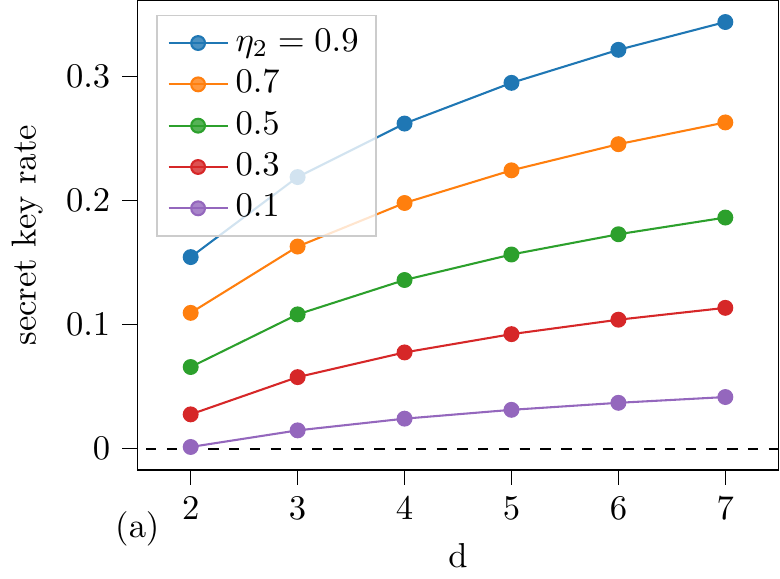}}
\subfloat{\includegraphics[]{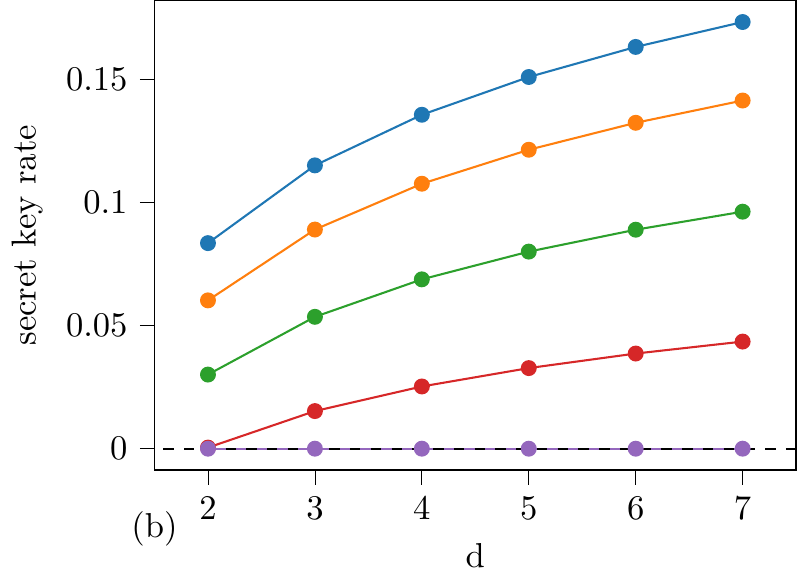}}
\caption{Secure key rates achievable using the $N=2$ case of the protocol. We simulate preparation of a 95\% fidelity W state where $\eta_{1}=0.9$, followed by creation and distribution of a qudit Bell state over a lossy channel with capture probability $\eta_{2}$. The resultant key rate is expressed in units of $R_{\pi}$, the rate at which emitters can be optically pumped. (b) Dephasing is introduced with $\gamma=0.01$ between each excitation pulse ($T_{2}/T_{\pi}=100$). Excessive losses cause the state generation time to become comparable to $T_{2}$, reducing the coherence of the final state and preventing the transmission of a secure key.}
\label{fig::bit_rate}
\end{figure*}

The generation of qudit GHZ states lends itself naturally to the sharing of secret information, protected by the laws of quantum mechanics. Restricting to the simpler case of 2 qudits, we can consider secret key rates achievable with this protocol via distribution of qudit Bell states. Secure key rates $K$ are calculated from measurement statistics of the generated photonic state in the computational and diagonal bases, following the methods in \cite{BaccoANetworks,Doda2021QuantumEntanglement}. In Figure~\ref{fig::bit_rate} we numerically investigate the total key rate $RK$ for Bell states transmitted over increasingly lossy channels, where again we target $\mathcal{F}_{W} = 95\%$. Higher dimensional encodings result in higher bit rates, and the enhancement increases for more lossy distribution channels. For very noisy channels ($\eta_{2} \lesssim 0.1$), the qubit approach fails to distribute a secure key, whereas the $d>2$ encodings succeed. In this regime, subspace encodings can also be used to boost the transmissable key rate~\cite{Doda2021QuantumEntanglement}. $\mathcal{F}_{W}$ is fixed at 95\% here for comparison, but will not give the optimal key rate in all scenarios - more loss requires higher fidelity W state preparation for optimal rates.

\setlength\figureheight{2.5in}
\setlength\figurewidth{0.45\textwidth}

\end{document}